# Near-IR photoluminescence from Si/Ge nanowire grown silicon wafers: effect of HF treatment

Seref Kalem,[1] Peter Werner,[2] Vadim Talalaev[2,3]

[1]National Research Institute of Electronics and Cryptology TÜBITAK-BILGEM, Kocaeli, Turkey;

[2]Department of Experimental Physics, Max-Planck-Institute, Halle (Saale), Germany;

[3]ZIK "SiLi-nano" Martin-Luther-Universität (Halle), [3]Karl-Freiherr-von-Fritsch-Str. 3 D - 06120 Halle, Germany

*Abstract*— We present the room temperature (RT) near-infrared (NIR) photoluminescence (PL) properties of Si/Ge nanowire (NW) grown silicon wafers which were treated by vapor of $HF:HNO_3$ chemical mixture. This treatment activates or enhances the PL intensity at NIR region ranging from 1000 nm to 1800 nm. The PL consists of a silicon band-edge emission and a broad composite band which was centered at around 1400-1600 nm. The treatment modifies the wafer surface particularly at defect sites particularly pits around NWs and NW surfaces by etching and oxidation of Si and Ge. This process can induce spatial confinement of carriers where band-to-band (BB) emission is the dominant property in Si capped strained Si/Ge NW grown wafers. Strong signals were observed at sub-band gap energies in Ge capped Si/Ge NW grown wafers. It was found that NIR PL is a competitive property between the Si BB transition and deep-level emission which is mainly attributable to Si related defects, Ge dots and strained Ge layers. The enhancement in BB and deep-level PL was discussed in terms of strain, oxygen related defects, dots formation and carrier confinement effects. The results demonstrate the effectiveness of this method in enhancing and tuning NIR PL properties for possible applications.

## I. Introduction

The development of a photonic Si based platform is fueled by the possibility of creating a quasi direct band gap Si and Ge light sources emitting at NIR region and particularly at telecommunication wavelengths, 1310 nm and 1550 nm. Short period Si/Ge superlattices (SL), quantum dots and nanowires can offer such possibilities through changes in their electronic band structure [1]-[6]. Strain induced modification of band structure and size reduction effects can favor direct band gap formation and related direct transitions [2]. Such changes would increase oscillator strength of optical transitions and thus leading to effective band gap engineering possibilities for applications in NIR [3]. However, number of defect related effects can take place due to increased amount of surface area and presence of strain fields due to oxidation. These effects can in turn mask or dominate intrinsic emissions originating from band gap engineered superlattice structures and quantum dots.

Despite the presence of sharp interfaces, the formation of Ge islands resulting in smeared interfaces and surface corrugation is typical for the Si/Ge multilayer growth [1]. Depending on growth conditions and structural parameters, the nature of RT PL from Si/Ge NW grown wafers could exhibit different properties. Ge nano clusters with sizes ranging from 2 nm to 8 nm in diameter exhibit PL peaks at 0.806 eV/1539 nm and 0.873 eV/1420 nm, whose positions are independent of cluster size [4]. RT PL peak at 1550 nm/0.8 eV in 6.5 monolayer (ML) Ge islands in Si was attributed to type-II recombination in Si/Ge quantum structures. However, the PL from Ge layers with coverages ranging from 1.7 to 6.8 ML was attributed to an evolution from quantum well-like to quantum dot-like emission [5]. It was claimed that these results were indicative of carrier localization within three dimensional (3D) confined states in a type-II band alignment. An enhanced band gap PL in strain symmetrical Si/Ge SL's was assigned to localized excitons bound to random potential fluctuations caused by variations in the strain distribution and layer thickness [6]. Same studies have shown that SL PL efficiency was stronger than corresponding alloy layer. These studies indicate the difficulty of associating RT PL bands with a clear origin. Strain can play an important role in the nature of the PL in NIR. One of the most important results of the strain and size effects in Si and Ge is the observation of enhanced transition associated with direct band gap material [7] [8]. However, mini-band formation through electronic band structure change from type-II (indirect) to type-I (direct) was proposed to explain the experimental results in strained Si/Ge superlattices [9].

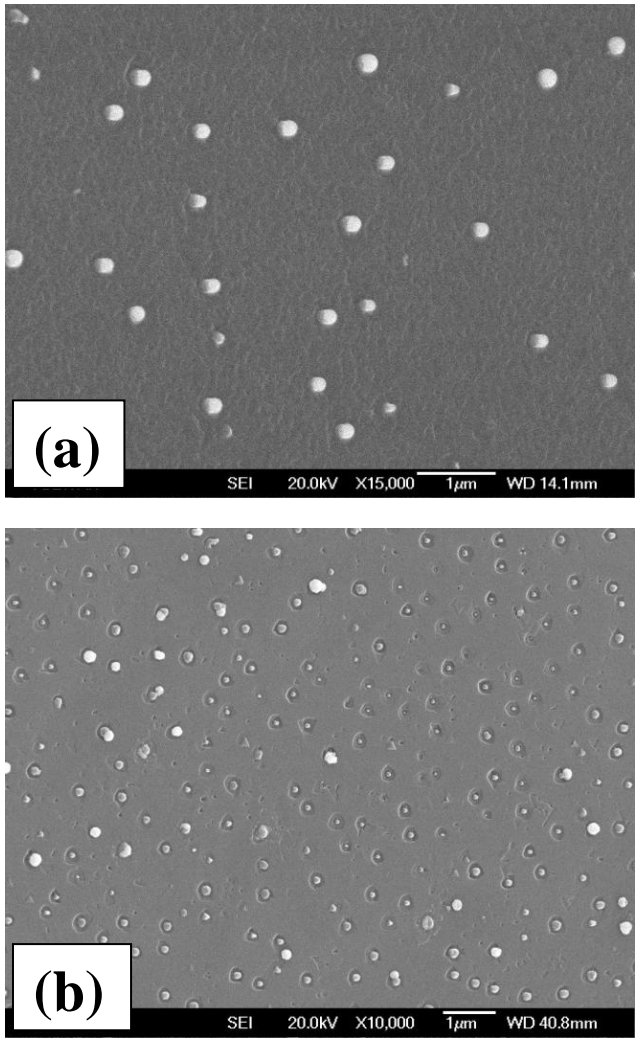

**Figure 1** SEM images of a Si(230nm)/Ge(2.5nm)/Si(40nm) NW grown on Si(200nm) buffer. Top image (a): as-grown, and the bottom image(b) after treated for 60 second(s) in acid vapor. The image after treatment indicates a porous like surface formation. Note that some of the NWs were removed from their locations following the treatment.

Creation of defects through dislocations and diffusion effects resulting in SiGe alloying introduce further complications in interpreting any PL spectrum of such quantum structures grown wafers. Reported emission energies in IR overlap with those of defects and impurities thus leading to speculations concerning the origin of the observed transitions. Additional complexity is due to the fact that PL excitation beam probes not only nanowires but also a large surface region on the wafer (typically, NWs occupy only about 10% of the wafer surface). This indicates a significant contribution to PL from the wafer surface and thus complicating the interpretation of the NW luminescence. The formation of relatively large pits or holes where the nanowires grow introduces further complications. Systematic characterization of optical and structural properties was carried out on Si/Ge single, multilayer and heterostructure NWs grown wafers in order to address the nature of the room temperature PL from Si/Ge NW grown Si wafers.

## II. Experimental

Si/Ge NWs were grown by molecular beam epitaxy (MBE) on Si wafers with <111> crystal orientation using Sb as a surfactant and a thin Au film with a nominal thickness of 2 nm [1]. The wafers consist of nanowires involving ultrathin Si/Ge multilayers (superlattices) or heterostructures with diameters of less than 100 nm. Thickness of individual layers is ranged from few monolayer (ML, where **$1ML = a/3^{1/2}$ ,a** being the lattice constant) to several tens of nanometer in the case of heterostructured nanowires. Each structure was insulated by a Si cap layer with a thickness ranging from 5 nm to about 40 nm finished by uncapped Ge layer. The details of the MBE growth were described elsewhere [1]. The NW grown wafers were subsequently treated by an exposure to vapor of $HF:HNO_3:H_2O$ based acid mixture and details are given elsewhere [10]. The main purpose for this treatment is to induce defects intentionally on NWs grown wafer surface in order to understand the nature of the NIR PL bands. Au was removed by a wet chemical process before the treatment involving $KI+I_2$ solution in order to avoid its possible perturbation. The PL was excited by a 488 nm line of an Ar+ laser at RT and the signal was collected by a liquid nitrogen cooled Ge photodetector between 800 nm-1800 nm. The results are combined with Fourier Transformed Infrared (FTIR) and energy dispersive spectroscopy (EDS) analysis using scanning electron microscopy (SEM) in an attempt to clarify the origin of the light emission. A typical SEM image is shown in Fig. 1(a) for a wafer consisting of a Si(230nm)/Ge(2.5nm)/Si(40nm) NW grown phosphorous doped Si(111) wafer with a resistivity of ~10-20 Ohm.cm. The wafer was treated in acid vapor for 1 minute resulting in a rough surface consisting of etch pits particularly the largest ones around the NWs as shown in Fig. 1(b). The acid vapor reacts on both the NW surface and particularly in the pits where defects are plenty favoring the reaction of HF and $HNO_3$ species on these sites. Note that the as-grown surface is not anymore a smooth single crystalline layer but resembles that of a multicrystalline structure due to a number of surface pits and depleted lattice at the bottom of each nanowire. On the other hand, the acid vapor transforms the surface to a porous-like Si or Ge depending on the cap layer. The acid vapor reacts with Si much more effectively as compared to Ge. Detailed TEM and SEM investigation of Si/Ge NW grown wafers studied her can be found elsewhere [11].

### III. Results and discussion

Acid vapor exposure can modify the Si/Ge nanowire grown Si surfaces in favor of $SiO_x$ and $GeO_x$ ($x \leq 2$) oxide formations involving probably Si and Ge nanocrystals or dots. The surface modification starts of preference at defect sites such as dislocation lines and particularly within the pits at around NWs and on the NW surfaces where chemical reactions are most favored. All this modification yields to an efficient NIR-infrared luminescence originating not only from Ge dots, but mainly from oxide related defects as well on the NWs grown wafer.

The Figure 2 summarizes the results of the PL measurements at 300K from Si/Ge heterostructure NWs grown Si(111) wafers, as compared with a reference wafer which

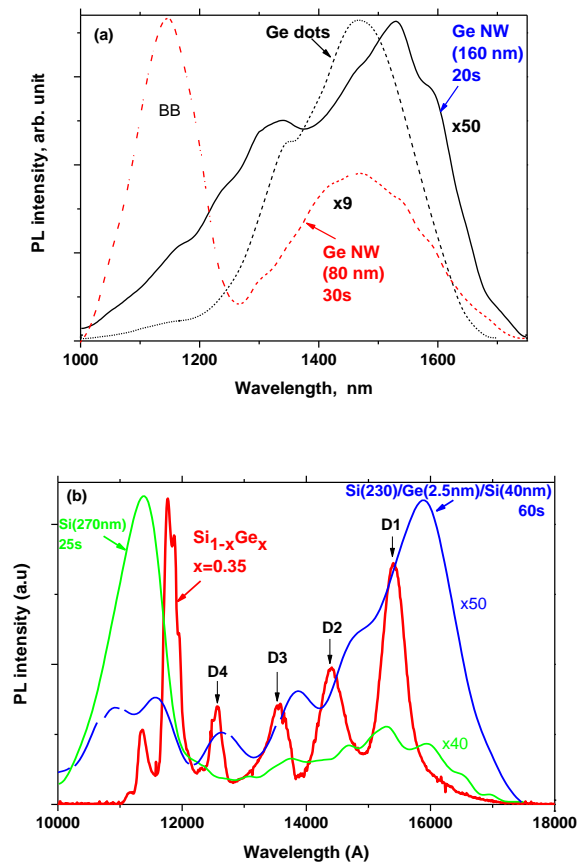

**FIGURE 2** (a) Room temperature PL from 80nm and 160 nm Ge NW grown on Si(100nm)/Ge(80nm) and Si(200nm)/Ge(100nm) buffer marked as dot-dash and solid lines, respectively. Au was removed from the first sample before exposure to vapor of HF:HNO$_3$ mixture for 30 second. The latter is an as-grown sample which was treated for 20 sec under acid vapor. The emissions are compared to a reference sample containing Ge dots of 200 nm wide and 5 nm high (dash line). Note that the dip at around 1360 nm is due to H$_2$O absorption; and (b) PL from a nanowire sample consisting of a single Ge layer of 2.5 nm which was treated for 60 second (solid line) and a 270 nm long Si NW grown on Si(8ML)/Ge(2ML) SL with a period 9 and an exposure time of 20 second. Symbol "x" denotes the amplitude factor relative to the intensity of Ge dot PL. Numbers in paranthesis are in nm. Also shown is a strain relaxed Si$_{1-x}$Ge$_x$ layer with dislocation lines D1, D2, D3 and D4 as measured at 4K.

contains Ge dots having dimensions of 5 nm high and 200 nm wide (dash line). The dip at around 1360 nm is due to water vapor absorption. This reference sample has a characteristic smooth PL peak with FWHM of 275 nm/151 meV at 1468 nm as shown in Fig. 2(a). Our samples exhibit variety of PL features at around Si band edge and further in near-IR depending on sample structure and treatment conditions. The thickness of the cap layer (when it is greater than 5 nm) has also significant effect on the emission, favoring the formation of dislocations and related emissions [1]. Such defects are preferred sites for etching by acid vapor consisting of HNO$_3$ as an oxidation and HF as an oxide etching agents [12].

The large composite emission band between 1200 nm – 1750 nm is structured and always persists in almost all of the luminescent structures. Despite some differences from sample to sample, it peaks at around 1550 nm. We show that Ge dots, strain and dislocations can contribute to this region. The dislocation-related PL in Si is known to consist of four lines as labeled D1 (0.8 eV/1550 nm), D2 (0.87 eV/1425nm), D3 (0.950 eV/1305nm) and D4 (1.0 eV/1240nm) [13]. The D1 dislocations located at around 1550 nm are the strongest as shown in Fig. 2(b). However, dislocation emission is usually observed at low temperatures. Here, the presence of D1-line should be attributed to a significant increase in the density of dislocations. Actually, earlier studies revealed an intense RT-PL and electroluminescence at around 1600 nm in plastically deformed Si and Si/Ge multilayers with dislocation densities of $\approx 10^9$ cm$^{-2}$ [13]. Thus, thus one can assume a likely contribution from dislocations to the composite band. Most recently, a RT sub-bandgap PL from Si containing oxide precipitates was observed at around 1600 nm and modeled by a transition between the bands of defect states associated with strained precipitates [14]. Samples exhibiting emissions at around 1600 nm, particularly Ge NW of 160 nm as shown in Fig. 2(a) could be of this type of nature. The most recent observation is about a deep-level PL at 0.79 eV/1570 nm due to dislocations and oxygen precipitates at 0.87 eV/1425 nm in multi-crystalline Si [15]. From the fact that our method induces oxidation at pits around depleted lattice and on the NW surfaces, it is reasonable to assume a contribution from oxidation related effects or oxygen related defects to the deep-level emission.

We have previously demonstrated that acid vapor exposure leads to effective oxygen diffusion into Si rods [16]. This leads us to think that the oxygen diffusion or precipitation and thus their strain field as in ref. [15] could be effective on our wafers. As suggested earlier [13] [14] [15] oxygen diffusion to dislocations can be at the origin of the enhancement of the IR emissions at around 1600 nm and 1100 nm. Sample dependent red-shift of the BB PL peak position supports the presence of a likely bi-axial compressive strain tending to red shift the band gap emission in Si [17]. It is reasonable to think that the strain plays an important role in determining the resultant surface properties.

The evidence for the strain or disorder can be obtained from the presence of Si-O coupled LO-TO modes through FTIR. The presence of these modes can readily be observed by transmission measurements as a function of the angle of incidence. Figure 3 shows the results of such measurements for the same sample, which is Si(230nm)/Ge(2.5nm)/Si(40nm) whiskers at 0°, 40° and 60°. The appearance and increasing oscillator strength of the strain related Si–O LO–TO splitting mode pair at 1255 cm-1 could be well the indication of the presence of stress or disorder in the film [18]. The enhanced oscillator strength of the mode at 1255 cm-1 as the incidence angle is changed, indicates the presence of a strain or disorder build-up in the pillar structure due to the formation of an oxide layer or oxidation [18].

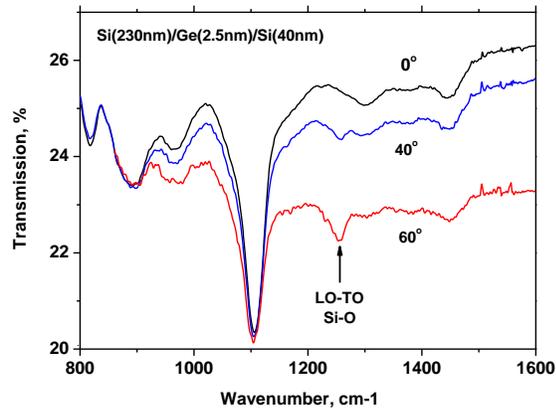

**Fig. 3.** FTIR spectrum indicating the presence of a strain related Si–O LO–TO coupling modes at 1255 cm−1 as appeared as a function of angle of incident light beam.

Treatment of Ge capped NWs grown wafers enhances IR emission between 1.0 µm and 1.75µm as shown in Fig.2(a). Note that no detectable emission could be observed from the as-grown samples. For 80 nm Ge NW grown on a Si(100nm)/Ge(80nm) heterostructured buffer layer, the PL measurement reveals a strong Si near band-edge emission at 1145 nm/1.083eV and a broad band at 1472 nm/0.842eV (dot-dash line). This wafer was treated for 30 sec after Au removal process in KI+$I_2$ solution. The emission should be an isolated or oxide encapsulated Ge dot-like emission by analogy with the reference Ge dot wafer since our process can induce oxidation on Ge [19]. Band edge emission (phonon-assissted band-to-band transition) at around 1150 nm/1.078 eV is originated from the buffer layer. Internal stress due to capping layer can induce dislocations which can cause this type of band edge emission. However, Si layer is underneath a 80 nm of Ge and only way of getting the strong emission from the Si sub-layer is through the pits, particularly those within the depleted regions [11] around NWs. The oxidation and thus oxygen diffusion should be particularly effective in these regions. For a thicker or longer Ge NW of 160nm grown on Si(200)/Ge(100) buffer and exposed to acid vapor for 20 sec, a broad emission band peaking at 1530 nm/0.81eV covers the whole IR region. Note that there is a weak Si band edge emission appearing as a shoulder from this sample as shown in Fig.2 (a). Relatively thick (160 nm) Ge layer suggests that the contribution of Ge related emission is highly significant. The broad nature of the band can be explained by the size distribution of Ge dots and strain. RT PL from ultra-pure Ge nanoparticles covered with GeOx (x<2) exhibits also a similar composite band peaking at 1575 nm and ranging from 1200 nm to 1700 nm. The shoulder at 1600 nm could be related to a residual tensile strain originating from the remaining Ge layer. Weak band edge transition should be attributed to thicker absorbing Ge layer and shorter exposure time.

PL properties of Si capped wafers are shown in Fig. 2(b). Relatively weak sub-gap emission and stronger near band emission is favored for a Si(270nm) NW grown on Si(200nm)/[Si(8ML)/Ge(2ML)]/Si(10nm) superlattice buffer of 9 periods which was treated for 25 sec as shown in Fig.2(b) (dash line). For this NW structure, strong Si band-edge transition at 1139 nm/1.089 eV. Here, it is likely that etching introduces minimal defects due to a better strain accommodation resulting in a mostly smooth crystalline surface favoring the stronger enhancement at band edge emission. This is in line with recent observations stating that the BB emission occurs when oxide precipitation is low [14]. Also shown in the same figure, the PL from a Si(230nm)/Ge(2.5nm)/Si(40nm) nanowire consisting of a single Ge well (exposed 60 second to acid vapor after the Au removal process). We observe a relatively weak near band emission at 1145 nm/1.083 eV and a strong PL enhancement at around 1550 nm (0.8 eV) consisting of a doublet at 1540 nm and 1590 nm, which is most likely due to an enhancement of oxygen related defects trapped around dislocations and a tensile strained Ge [15]. It is likely that the capping layer contains a large number of pits and faults, inducing sever etching reactions resulting in heavy oxidation and oxygen diffusion and thus leading to strained oxygen precipitation. Figure 4 compares the PL spectra of as-grown and the treated samples indicating a significant enhancement of the PL. The first wafer consists of Si NWs of 270 nm which was treated for 20s. The enhancement of luminescent intensity at the band-edge is particularly significant. The increase of PL intensity here is due to spatial confinement of carriers as a result of oxidation. The insert in the same figure compares a treated (for 60s) and as-grown wafers of Si(230nm)/Ge(2.5nm)/Si(40nm) NWs. The enhancement was observed at around 1400-1600 nm. This type of enhancement could be observed when lateral dimensions of Si/Ge/Si single quantum wells were reduced [19]. Our treatment can induce such localization enhancement by selectively etching Si rich regions and thus isolating Ge dots.

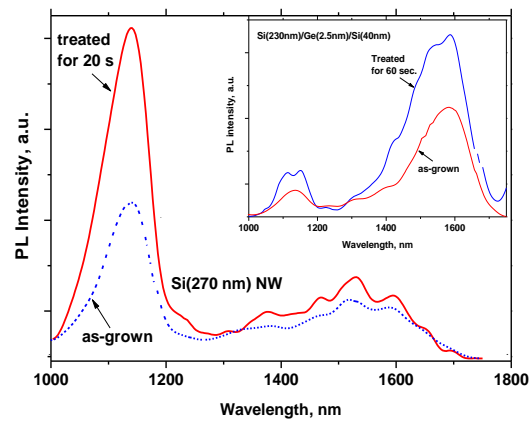

**Fig. 4** Photoluminescence spectra of as-grown and treated wafer which was exposed to acid vapor for 20 second. The wafer contains Si NWs of 270 nm long. The insert compares also an as-grown and treated wafer of Si(230nm)/Ge(2.5nm)/Si(40nm) NW.

The acid vapor etching can form Ge islands and isolate dots by oxidation (formation of α-$GeO_2$) [20] and thus inducing more effective absorption of light as well as confinement of carriers within Ge dots. The formation of oxide encapsulation was confirmed by FTIR experiments through the presence of infrared active Si-O and Ge-O stretching modes at 1107 cm-1 as interstitial oxygen and around 900 cm-1 possible signature of Ge oxide precipitate, respectively. When compared an as-grown and treated (for 20 sec) Si(360nm)/Ge(20nm) NW sample, the acid vapor exposure only induces changes at around 1107 cm-1 through the broadening of Si-O-Si stretching band and appearance of a shoulder at 1060 cm-1. We attribute this change to mixed vibrational modes, which is Si-O-Ge. By analogy with previous experimental results [21] and theoretical estimations [2], confinement energies ranging from 1700 nm to 1000 nm would correspond to Ge quantum dots of about 2 to 6 nm. The enhancement and broad nature of the IR emission band would suggest an effective oxide encapsulation around Ge dots, thus supporting the presence of a direct radiative recombination process in Ge quantum dots in addition to defect emission originating from Si.

Energy dispersive spectrum (EDS) confirms also the observations in FTIR spectrum as shown in Fig. 6. EDS is measured at the surface of an Si/Ge NW grown wafer, consisting of Si(360)/Ge(20) NW which was grown on Si(100) buffer followed by 2 nm of Au layer. It is an as-grown wafer which was treated in acid vapor for 20 seconds. Despite the signal includes the bulk contribution, EDS is a useful tool to determine the surface oxidation of the sample. The figure reveals the presence of a relatively strong oxygen peak.

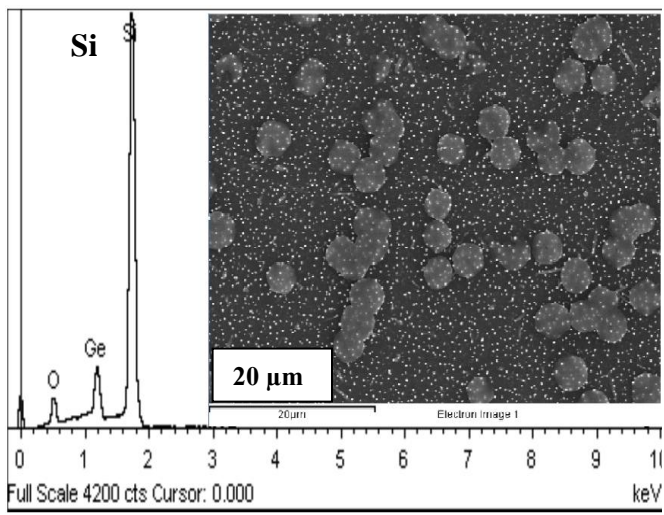

**Figure 5** Typical energy dispersive spectrum as measured at the surface of a Si/Ge nanowire grown wafer, where Si(360nm)/Ge(20nm) NW was grown on Si(100nm) buffer. It is an as-grown wafer which was treated in acid vapor for 20 seconds. EDS was measured at the round shaped oxide precipitates.

The probe located on white circular spots at the inserted figure measures Oxygen, Si and Ge concentrations of 32.92, 62.23 and 4.85 atomic percent, respectively. Round shaped regions are therefore attributed to oxide precipitates. The IR PL for this sample doesn't exhibit any BB emission but an emission around 1540 nm. White dots are gold but EDS probe indicates they are composed of O, Si, Ge and Au having concentrations of about 20, 75, 3 and 2, respectively.

The strong band edge emission at around 1130 nm/1.1 eV in Fig. 2 could be the result of a spatial confinement of carriers in Si crystallites [23] leading to an efficient luminescence. A RT scanning PL revealed a strong near band edge emission at 1.09 eV/1138 nm in microcrystalline silicon (μc-Si) [24]. Resemblance of the NW grown surface to multicrystalline structure as shown in Fig.1 can be supportive of this finding. However, we observe shifted peak positions ranging from 1130 nm to 1150 nm (a shift of about 20 meV) depending on the sample structure and the treatment indicating that other effects could be well involved. A red shifted near-band-edge emission at 1154 nm was reported to be due to dislocations in Si implanted with Boron and Si ions [24]. The same work proposed a model based on one-dimensional (1-D) energy bands associated with the strain field of dislocations. However, the possibility of alloying of Ge with Si as a result of intermixing can induce similar shifts at the band edge in Si/Ge NW grown wafers. Impurities can play an important role in low energy band-edge emission. They can disrupt local translational symmetry and thus relaxing the radiative selection rules. This effect can enhance the relative strength of impurity transitions at the band-edge of silicon. However, the observed BB luminescence peak ranges from 1130 nm up to 1150 nm depending on the duration of the acid vapor exposure. A recent report on epitaxial chemical vapor grown Si NW considered the role of the many-body interactions and spatial confinement of carriers in explaining the smaller recombination energy and line broadening effects as compared to the bulk Si [25]. The oxidation can indeed enhances spatial confinement of carriers thus leading to an enhanced emission. The BB peak position in our wafers increases from 1130 nm to 1150 nm for the treatment duration of 5s and 60s, respectively. But further studies are required to distinguish between these effects and those related to strain.

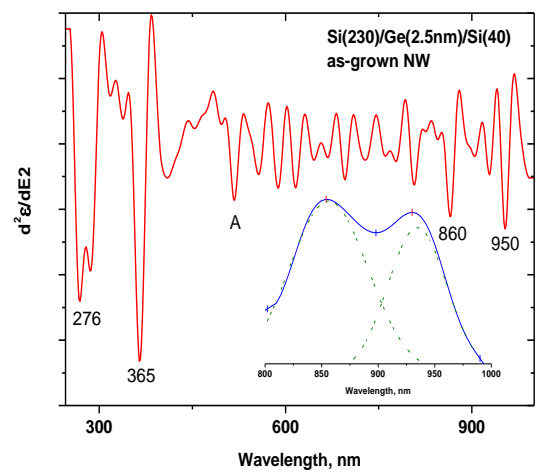

**Figure 6** Second derivative of the dielectric function of an as-grown NW of Si(230)/Ge(2.5nm)/Si(40) indicating critical point energies of electronic transitions where A is an experimental artifact.

Treated Si/Ge nanowires grown wafers exhibit also an intense photoluminescence in the region of λ < 1000 nm which can be attributed to confined carriers in smaller Ge dots [26]. Whereas the higher energy bands (610 nm and 700 nm) are probably due to the confinement in smaller Si or Ge dots [27] embedded in a Si oxide layer as the cap layer which was formed during the acid vapor exposure. The related electronic transitions as measured by ellispsometry were shown in Figure 6 where possible contributions to PL are indicated by arrows. The RT PL from the same sample is also shown as an insert in Fig.6 in which the emission peaks at around 900 nm could be corresponding to critical points observed from the spectroscopic ellipsometry measurements. One of the earlier reports claimed the contribution of these visible emissions to deep-level bands in IR as a second order emission in Si NWs [28]. But this hypothesis needs to be confirmed by further studies combining visible PL and spectroscopic ellipsometry studies.

## IV. Conclusion

The treatment of Si/Ge NW grown Si wafers by vapor of $HF:HNO_3$ chemical mixture activates or enhances photoluminescence in near infrared region. The results indicate that treatment induces significant changes in the wafer's surface. The changes consist of oxidation and oxidation induced strain and confinement of carriers in Si and Ge. The modification results in activation and enhancement of the PL originating from the Si band-edge and deep-levels up to 1800 nm. The band edge PL is strongly enhanced and red-shifted with the duration of the treatment. The red-shift was found to be attributable to strain related effects due to surface oxidation as a result of acid vapor exposure. The treatment induces also spatial confinement due to crystal size reduction as a result of oxidation or oxide encapsulation of crystals. Defects and Ge dots are the main contributors to deep-level emission between 1200 nm and 1800 nm. The enhancement in this region could be attributed to an increase in the number of Si related defects, strained Ge layers and lateral confinement of carriers in Ge as a result of crystal size reduction due to oxidation. The heavy oxidation at pits around NWs is effective in confining Si and Ge layers further reducing their crystal sizes. The treatment also enables more efficient collection of emission from the buffer layers through the pits thus contributing to deep-level emission from Si related defects. Oxide formation and oxygen diffusion around surface defects and particularly those around NWs can have a significant contribution to the PL enhancement in sub-gap region at around 1590 nm. However, the band-edge emission dominates if the cap layer is silicon. For Ge capped wafers, the contribution to sub-gap emission is enhanced by lateral reduction in dimensions due to etching and oxide formation around Ge rich regions. The results indicate that there is a significant contribution from defects in Si. Dislocations particularly D1 in relaxed layers are likely contributing to the emission at around 1200-1600 nm. Therefore, any PL analysis should be made carefully taking into account Si related defects and their contribution to deep level emission (composite band). The results suggest that IR emission lines can be controlled by choosing an appropriate structural parameter followed by defect engineering through acid vapor treatment. Further studies are required for detailed description of these defects and the nature of the deep-level PL.

ACKNOWLEDGMENT

This work was supported by TUBITAK (TBAG) bilateral program under contract No. 107T624, and BMBF German Federal Ministry of Education and Research (Grant No: 03Z2HN12).

REFERENCES

1. N.D. Zakharov VG Talalaev, P Werner, AA Tonkikh and GE Cirlin, "Room temperature light emssion from a highly strained Si/Ge superlattice" Appl. Phys. Lett. **83**, 3084-3086(2003)
2. Y M Niquet, G. Allan, C.Delerue and M. Lanoo, Appl. Phys. Let. **77**, 1182(2000)
3. Y-H Kuo and Y-S Li, Applied Physics Letts. **94,** 121101(2009)
4. G.E. Cirlin et al. Physica E **17**, 131(2003)
5. M W Dashiell et al. Applied Phys. Letts. **80,** 1279(2002)
6. U Menczigar, G. Abstreiter, J.Olajos, H.Grimmeiss, H.Kibbel, H.Presting and E.Kasper Phys. Rev. B**47**, 4099(1993)
7. W.D.A.M. de Boer et al, Nature Nanotechnology 5, 878(2010)
8. J Liu et al. Opt. Express **15**, 11272(2007)
9. VG Talalaev et al. Nanoscale Research Letts **1,** 137(2006)
10. S Kalem and O Yavuzcetin, Optics Express **6,** 7(2000)
11. P. Werner, et al., Int. J. Mat. Res. 97, 7(2006)
12. E.S. Kooij et al., Electrochem. & Solid State Letters 2, 178(1999)
13. E Ö Sveinbjörnsson and J.Weber Thin Solid Films **294**, 201(1997).
14. K. Bothe, R.J. Folster and J.D. Murphy, Appl. Physics Letts 101, 032107 (2012)
15. M.Tajima et al., J. Appl. Phys. 111, 113523(2012)
16. S Kalem, P Werner, B Nilsson, V Talalaev, M Hagberg, O Arthursson and U Sodervall, Nanotechnology **20,** 445303(2009)
17. H. Cui et al., Nano Letters 8, 2731(2008)
18. Kirk C T 1988 *Phys. Rev.* B **38,** 1255
19. R. Chivas et al., Optical Material 33, 1829(2011)
20. S Kalem, Ö Arthursson and I Romandic Thin Solid Films **518,** 2377(2010)
21. S.H. Choi et al., J. Korean Phys. Soc. 42, S120 (2003)
22. K.L. Kavanagh, Semicond. Sci. Technol. **25,** 024006(2010)
23. L Tsybeskov, K L Moore, D.G Hall, P M Fauchet, Physical Review B **54**, R8361(1996).
24. I Tarasov et al. Mat. Sci. & Engin. B**71,** 51(2000)
25. O Demichel et al., Physica E 41, 963(2009)
26. D.J Stowe, S.A. Galloway, S. Senkader, K Mallik, J. Falster, P.R. Wilshaw, Physica B **340–342**, 710(2003)
27. K.W. Sun et al., Physica E28, 525 (2005).
28. G. Jia et al., Phys. Stat. Sol. (a) 203, R55(2006)